\documentclass[twocolumn,amsmath,amssymb,prd,widetext]{revtex4}
\usepackage{bm,epsfig}
\usepackage{amsmath}
\usepackage{epsf}
\usepackage{color}

\newcommand{\beq}{\begin{equation}}

\newcommand{\eeq}{\end{equation}}
\newcommand{\beqa}{\begin{eqnarray}}
\newcommand{\eeqa}{\end{eqnarray}}

\newcommand{\Da}{{\cal D}}

\newcommand{\btheta}{\bm{\theta}}
\newcommand{\zlens}{z_{\rm L}}
\newcommand{\zsource}{z_{\rm S}}

\begin{document}

\title{CMB Cluster Lensing: Cosmography with the Longest
Lever Arm}
\author{Wayne Hu$^{1}$, Daniel E. Holz$^{2}$,  Chris Vale$^{3}$}
\affiliation{
$^{1}$Kavli Institute for Cosmological Physics, Department of Astronomy and Astrophysics,
and Enrico Fermi Institute,
University of Chicago, Chicago IL 60637 \\
$^{2}$Theoretical Division, Los Alamos National Laboratory, Los Alamos, NM 87545\\
$^{3}$Particle Astrophysics Center, Fermilab, P.O. Box 500, Batavia, IL 60510
}

\begin{abstract}
\baselineskip 11pt 
We discuss combining gravitational lensing of galaxies  and the cosmic microwave
background (CMB)  by clusters to measure cosmographic distance ratios, and hence dark energy parameters.
Advantages to using the CMB as the second source plane, instead of galaxies, 
include:
a well-determined source distance, a longer lever arm for distance ratios, typically 
up to an order of magnitude
higher sensitivity to dark energy parameters, 
and a decreased sensitivity to photometric redshift accuracy of the lens
and galaxy sources.
Disadvantages include: higher statistical errors, potential systematic errors,
and the need for disparate surveys that overlap
on the sky.   Ongoing and planned surveys, such as the South Pole
Telescope in conjunction with the Dark Energy Survey, can potentially reach the statistical
sensitivity to make interesting consistency tests of the standard cosmological constant model. 
Future measurements that reach $1\%$ or better precision in the convergences 
 can provide sharp tests for future supernovae
distance measurements.
\end{abstract}
\maketitle

\section{Introduction}

Gravitational lensing depends 
on the distances between the observer, lens, and source.   These distances provide geometric
measurements of the expansion
history of the universe in much the same way as distant
supernovae.
Measurements of the same
lens with multiple source planes can be used to construct distance ratio estimates
that are, in principle, independent of the mass distribution (e.g.~\cite{LinPie98, GauForMel00, GolKneSou02, Ser02}).
In the weak lensing regime, where measurement and projection errors on individual
lenses are large, these ratios can be measured statistically by stacking multiple lenses
\cite{JaiTay03,BerJai04}
or equivalently by measuring correlation functions \cite{HuJai03,ZhaHuiSte05,Johetal05}.

In this {\it Brief Report} we examine the use of recently developed 
CMB cluster mass reconstruction techniques \cite{HuDeDVal06} (see also
\cite{SelZal00,Hu01b, HirSel02,ValAmbWhi04,MatBarMenMos05,LewKin06}) for measuring distance ratios.
We discuss the benefits and drawbacks to using the CMB as a lensing source plane,
and assess the impact future surveys may have on dark energy parameter
measurements.
For illustrative purposes, we describe the cosmology with the following
parameters (values in square brackets denote our adopted fiducial choices).
On the low redshift side:  the dark energy 
density in units of the critical density $\Omega_{\rm DE}\,[=0.76]$, dark energy
equation of state $w(a)=w_0 +(1-w_a)a\,[=-1]$, and spatial curvature $\Omega_K\,[=0]$.
On the high redshift side: matter density
$\Omega_m h^2\,[=0.128]$, baryon density $\Omega_b h^2\,[=0.0223]$, optical
depth $\tau\,[=0.092]$, tilt $n\,[=0.958]$, and scalar amplitude $\delta_\zeta\,[=4.52 \times 10^{-5}]$
at $k=0.05$Mpc$^{-1}$.  

\section{Cosmographic Distances}
\label{sec:likelihood}

Gravitational lensing of galaxy images or the CMB at a redshift $z_{\rm  S}$ 
by an object at redshift $\zlens$
with comoving surface mass density
$\Sigma$  can be phrased in terms of the convergence
\begin{equation}
\kappa({\btheta}, \zlens,\zsource) = {4\pi G} {\Da_{\rm L}}{\Da_{\rm  LS} \over {\Da_{\rm S}}} (1+\zlens) \Sigma(\Da_{\rm L}{\btheta},\zlens)\,,
\end{equation}
where ${\Da_{\rm L}}$, $\Da_{\rm S}$, and $\Da_{\rm  LS}$ are the comoving angular diameter 
distances from observer to lens, observer to source, and lens to source respectively. 
Here $\btheta$ denotes the angular position on the sky.

 In the idealization of
perfect measurements at all angular positions and all the lensing being
generated by a single lensing plane, the ratio of the measured convergence
for two different source planes,  $\zsource$ for the galaxies  and $z_*$ for the
CMB, depends only on the distance
ratio \cite{LinPie98, GauForMel00, GolKneSou02, Ser02,JaiTay03}:
\begin{equation}
R(\zlens,\zsource) \equiv {\kappa(\btheta,\zlens,\zsource) \over \kappa(\btheta,\zlens,z_*)}
  = {\Da_{\rm LS} \over \Da_{\rm L*} } 
     {\Da_{*} \over {\Da_{\rm  S}}} \,.
     \label{eqn:Rdef}
  \end{equation}
One virtue of using the CMB for the second source plane is that $\Da_*$ is measured
to high precision from the positions of the acoustic peaks.  For example, in the projections for
the Planck satellite (see below), the fractional error in distance $\sigma(\ln \Da_*)=0.002$.

\begin{figure}[t]\begin{centering}
\includegraphics[width=0.9\columnwidth]{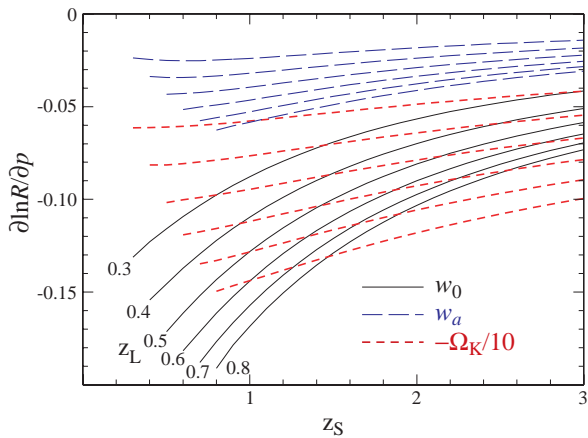}
\caption{Sensitivity to cosmological parameters of the convergence ratio $R$ between the CMB last scattering surface and
a galaxy source redshift $\zsource$.  Utilizing the CMB as a source
plane can boost the sensitivity to parameters typically by up to an order of magnitude.}
\label{fig:efficiency}
\end{centering}\end{figure}

The second virtue of using the CMB as a source plane is that the large separation between
it and
typical galaxy source planes boosts the sensitivity of the ratio to cosmological parameters.  
In Fig.~\ref{fig:efficiency} we show the sensitivity of $R$ to $w_0$, $w_a$, and $\Omega_K$ 
assuming that the high redshift parameters and $\Da_*(\Omega_{\rm DE})$ are fixed.  A percent level determination
of $R$ with $\zlens < 1$ and $\zsource\sim 1$ would provide interesting
constraints on the dark energy and the curvature.
Contrast this with the sensitivity of the convergence ratio between two galaxy source
planes ($z_1$, $z_2$)
\begin{equation}
G(\zlens,z_1,z_2) = {\kappa({\btheta},\zlens,z_1) \over \kappa({\btheta},\zlens,z_2) }
= {R(\zlens,z_1) \over R(\zlens,z_2)} \,,
\end{equation}    
which is typically an order of magnitude less since it requires a measurement
of the much smaller  change in $R$ with galaxy source redshift 
\begin{align}
{\partial \ln G \over \partial p} & = {\partial \ln R \over \partial p}\Big|^{\zsource=z_1}_{\zsource=z_2} \,.
\end{align}

The  insensitivity of $R$ to galaxy source redshifts around $\zsource\sim 1$ also implies
that the requirements on measuring galaxy photometric redshifts is much less stringent than
for $G$ ({\it cf}. \cite{BerJai04}).
For example, the sensitivity of the ratio to redshift around $\zlens=0.7$ and $\zsource=1$ is
    \begin{eqnarray}
  { \partial\ln R \over \partial \zlens}\Big|_{\zlens=0.7} = -3.4\,, \qquad 
  {\partial \ln R \over \partial \zsource}\Big|_{\zsource=1.0} = 2.3 \,,
  \label{eqn:zsens}
  \end{eqnarray}
so that a measurement of $R$ to a few percent requires photometric redshifts
that are unbiased to $~1\%$.  Furthermore, given the weak dependence of $R$ on 
redshift, high precision in the photometric redshifts of individual galaxies is not required.

\begin{figure}[tb]\begin{centering}
\includegraphics[width=0.9\columnwidth]{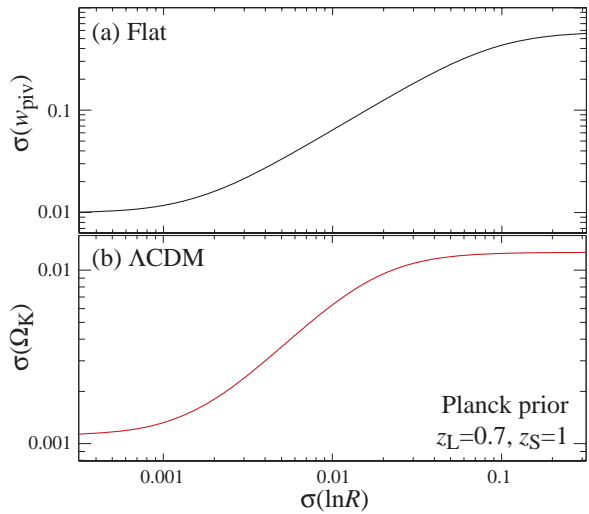}
\caption{Impact on parameter errors given Planck CMB power spectrum prior
for (a) the equation of state at the best constrained redshift $w_{\rm piv}$ in a flat cosmology and
(b) the spatial curvature in a cosmological constant ($\Lambda$CDM) cosmology.   Lens and
source redshifts here are $\zlens=0.7$ and $\zsource=1$.}
\label{fig:planck}
\end{centering}\end{figure}
  
  \section{Forecasts}

In practice, due to measurement errors and projection effects, cosmographic distances for
individual objects like clusters of galaxies are too noisy to be useful.   Instead multiple clusters
can be stacked in order to measure a cluster-mass correlation function or average profile
\cite{JaiTay03,BerJai04,HuJai03}.
  Projection
effects from mass along the line of sight that is not associated with the cluster, which can introduce $\sim 30\%$ scatter in the mass estimates of individual
clusters (e.g.~\cite{Metetal99}), averages away in this measurement \cite{Johetal05}.
Given the weak sensitivity of $R$ to the lens and source redshift distribution compared
with expected photometric redshift measurements, we can treat this statistical measurement
as providing the average $\kappa$ at the median lens and source redshifts for forecasting purposes.
Furthermore, dividing up the distribution into multiple lens and source planes 
does not provide much leverage for parameter estimation (see Fig.~\ref{fig:efficiency}).
For simplicity we will thus
treat each pair separately.

With upcoming weak lensing surveys such as the Dark Energy Survey (DES),
the expected statistical errors on $R$ will be dominated by the CMB measurements.
Hu et al. \cite{HuDeDVal06} estimate that the statistical errors for clusters above a mass of
$10^{14.2} h^{-1} M_\odot$ at $z\approx 0.7$ equate to
a $\sim 10\%$ rms error for $\kappa$ at the $\sim 1'$ scale radius 
per 1000 clusters.  This assumes a survey with $10\mu$K$'$ instrument noise,
comparable to the statistical sensitivity of the
ongoing South Pole Telescope (SPT) experiment,  but with no foreground contamination from
the cluster.   With an expected yield of
$\sim 10^4$ clusters, the statistical precision can 
reach $\sim 3\%$ in $\kappa$ or $R$.   Furthermore, with longer integration times an
experiment can improve on these numbers by a factor of 3--4 as the 
sample variance limit of temperature based estimators  is reached.   Lower
mass objects such as the luminous red galaxies selected in DES can also serve 
as lenses.  Finally, polarization measurements
with sensitivity in the $\sim 1\mbox{--}3\mu$K$'$ range, comparable to SPTpol, can
provide the means for achieving further improvements and checks for
systematic errors \cite{HuOka01,HuDeDVal06}.

  Since an actual measurement
will likely be dominated by systematic errors and foregrounds, we will phrase our forecasts in
an experiment-independent manner.  Given a measurement of $R$ to a certain fractional
precision $\sigma(\ln R)$, the information on a set of parameters $p_i$ is quantified
by the Fisher matrix
  \begin{equation}
  F^{R}_{ij} = {\partial \ln R \over \partial p_i} {1 \over \sigma^2(\ln R)}{\partial \ln R \over \partial p_j} \,.
  \end{equation}
  The inverse of the Fisher matrix provides an estimate of the covariance matrix between the parameters such that $\sigma(p_{i}) \approx [{\bf F}^{-1}]_{ii}$.
  Given multiple cosmological parameters and a single $R$, the Fisher matrix is degenerate
  and only one direction in the parameter space can be constrained.  While multiple lens and source
  planes provides some opportunity to break the degeneracies, it is more useful to examine 
  how a measurement of $R$ will complement other measurements in the future.
  
  We first combine the measurement of $R$ with those of the CMB power spectrum
  expected from the Planck satellite.   These measurements are also required to fix the distance
  to last scattering $\Da_*$ in Eq.~(\ref{eqn:Rdef}).   Details for the construction of the
  Planck Fisher matrix are given in \cite{HuHutSmi06}; we assume 
  80\% sky and 
3 channels: FHWM $5.0'$ 
with temperature noise $\Delta_T = 51\mu$K$'$ and polarization noise
$\Delta_P = 135\mu$K$'$ ; $7.1'$ with $\Delta_T=43\mu$K$'$, $\Delta_P=78\mu$K$'$;
and $9.2'$ with $\Delta_T = 51\mu$K$'$, $\Delta_P =\infty$.

Fig.~\ref{fig:planck} shows the errors in the equation of state at the
best measured redshift $w_{\rm piv}$ in a flat cosmology (see e.g.~\cite{HuJai03})
and $\Omega_K$
in a $w=-1$ $\Lambda$CDM cosmology as a function of $\sigma(\ln R)$ for
$\zlens=0.7$ and $\zsource=1.0$.  These two parameters benchmark how well
the standard flat $\Lambda$CDM cosmology can be tested or excluded.
Note that improvements in parameter estimation begin with 10\% measurements
of $R$.  Strong consistency checks are possible with $1\%$ measurements.  
To utilize $0.1\%$ measurements, improvements beyond Planck on the high
redshift parameters will be required. 

Other choices of lens and source redshifts in this range provide similar 
results.   Increasing the source redshift to $\zsource=1.2$ degrades the errors on $w_{\rm piv}$
by $12\%$.  Decreasing the lens redshift to $\zlens=0.6$ with $\zsource=1$ 
degrades them by $9\%$.  Increasing the lens redshift to $\zlens=0.8$ with $\zsource=1$
improves the measurement of $w_{\rm piv}$ by $6\%$.

\begin{figure}[tb]\begin{centering}
\includegraphics[width=0.9\columnwidth]{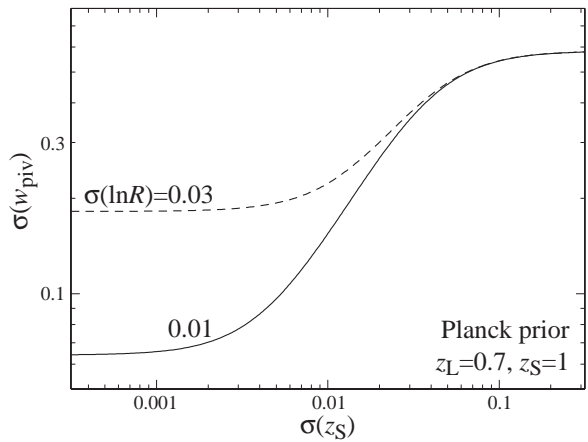}
\caption{Requirements on photometric redshift accuracy imposed by demanding that
$w_{\rm piv}$ measurements not degrade substantially  for 1\% and 3\% measurements 
of $R$ at $\zlens=0.7$ and $\zsource=1$.  A flat $w_0\mbox{--}w_a$ cosmology is assumed.}
\label{fig:photozs}
\end{centering}\end{figure}

In Fig.~\ref{fig:planck} we assumed that the redshifts of lens and source were perfectly 
determined.   To assess the precision with which they need to be measured we
add them as parameters in the Fisher matrix.  In Fig.~\ref{fig:photozs} we show
the degradation of errors on $w_{\rm piv}$ with imperfect knowledge of the mean 
of the source photometric redshifts.    To fully utilize $1\%$ measurements of $R$, one
requires a redshift accuracy of $\sigma(\zsource)\sim 0.003$
 whereas $3\%$ requires only $\sim 0.01$ accuracy.
 Source redshifts are more problematic than lens redshifts due
to the large number density of sources and their higher redshift.  Furthermore,
with cluster lenses, multiple red galaxy cluster members can be used to estimate
the redshifts.  Nonetheless, sensitivity to lens redshift measurements can also be inferred
from Fig.~\ref{fig:planck} by rescaling with the ratio of derivatives in Eq.~(\ref{eqn:zsens}).

 Finally we assess how well CMB lensing measurements complement the combination of future
supernovae (SNe) distance measures and Planck.   For the SNe, we assume
a sensitivity comparable to the proposed SNAP satellite and adopt the prescription
described in \cite{HuHutSmi06}; we take $2800$
SNe 
distributed in redshift out to $z=1.7$ according to \cite{SNAP}, 
300 local supernovae uniformly distributed in the $z=0.03\mbox{--}0.08$ range, 
statistical magnitude errors of $\sigma_m = 0.15$ per SN, and a systematic floor
 of  $\sigma_{\rm sys} = 0.02\,(1+z)/2.7$ per
$\Delta z=0.1$.

\begin{figure}[tb]\begin{centering}
\includegraphics[width=0.9\columnwidth]{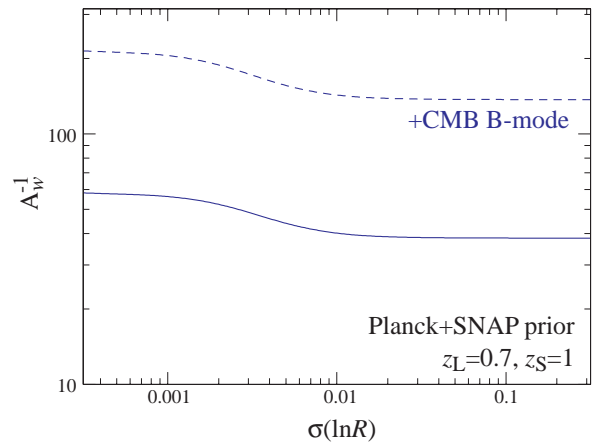}
\caption{Impact on  the inverse area statistic $A_w^{-1}$ of the error ellipse for the
equation of state parameters $w_0$, $w_a$ given SNAP supernova  and Planck priors with 
curvature marginalized.
Solid line: $R$ measurement only.  Dashed lined: including CMB $B$-mode power spectrum
measurements of gravitational lensing comparable to SPTpol.}
\label{fig:fom}
\end{centering}\end{figure}

In Fig.~\ref{fig:fom} (solid curve) we show the impact on the area statistic
of the $w$ error ellipse,  $A_w = \sigma(w_{\rm piv})\sigma(w_a)$ \citep{hut_turner},
with $\Omega_{K}$ marginalized.
Until errors reach below $1\%$, $R$ measurements do not provide
significant parameter error
improvements.  Nonetheless, $R$ measurements in the $\sim 1\%$ range 
do provide strong, {purely geometrical} consistency
tests on supernovae measurements.

 CMB lensing can improve $A_w$ more significantly but
the leverage comes mainly from lensing by large-scale structure.   In the dashed lines
we show the further improvement by including the forecasted constraints from $B$-mode
polarization power spectrum measurements by SPTpol \cite{HuHutSmi06}.   
With both sets of lensing information combined, the improvement in $A_w$ can ultimately
reach a factor of $5.5$.

\section{Discussion}
\label{sec:discussion}

We have assessed the potential of joint cluster gravitational lensing measurements 
from the CMB and weak {galaxy} lensing surveys for determining distance ratios.  These distance ratios
are in turn sensitive to dark energy parameters and can be used to 
test the flat $\Lambda$CDM model.   Benefits to using the CMB as a source
plane include a well-determined source distance, a longer lever arm and thus higher
signal, and a decreased sensitivity to photometric redshift errors of the lens
and galaxy sources.

We show that  if convergence ratios can be measured at percent level accuracy,
the dark energy equation of state can be measured to $\sim 6\%$ when combined with
CMB information from Planck in a flat universe.   Such a measurement would provide an interesting
consistency check on inferences from supernovae distance measures. 
Statistical errors of a few percent should be achievable
with existing and planned cluster surveys, such as the SPT in combination with DES.
However, the measurement will likely be limited by systematic errors, mainly on the
CMB side.  Minimum
requirements include a high signal-to-noise CMB map of at least 10$'$ resolution  that is cleaned of the thermal
Sunyaev-Zel'dovich effect in clusters \cite{ValHu07}.  
Although a full assessment is beyond the scope of this {\it Brief
Report}, we have
shown that the combination of galaxy and CMB source planes have the potential to provide strong
constraints on cosmological distance ratios, and thus make interesting contributions
to our knowledge of the dark energy.

\noindent{Acknowledgments:} We thank Eric Linder, Michael Mortonson, and Amol Upadhye for useful conversations.
W.H. was supported by the DOE, the Packard Foundation, and the KICP
under NSF PHY-0114422.  D.E.H. acknowledges support from a Richard Feynman Fellowship at LANL.
C.V. was supported by the U.S. Department of Energy and by NASA grant NAG5-10842.

\bibliography{HuHol07}

\begin{thebibliography}{22}
\expandafter\ifx\csname natexlab\endcsname\relax\def\natexlab#1{#1}\fi
\expandafter\ifx\csname bibnamefont\endcsname\relax
  \def\bibnamefont#1{#1}\fi
\expandafter\ifx\csname bibfnamefont\endcsname\relax
  \def\bibfnamefont#1{#1}\fi
\expandafter\ifx\csname citenamefont\endcsname\relax
  \def\citenamefont#1{#1}\fi
\expandafter\ifx\csname url\endcsname\relax
  \def\url#1{\texttt{#1}}\fi
\expandafter\ifx\csname urlprefix\endcsname\relax\def\urlprefix{URL }\fi
\providecommand{\bibinfo}[2]{#2}
\providecommand{\eprint}[2][]{\url{#2}}

\bibitem[{\citenamefont{{Link} and {Pierce}}(1998)}]{LinPie98}
\bibinfo{author}{\bibfnamefont{R.}~\bibnamefont{{Link}}} \bibnamefont{and}
  \bibinfo{author}{\bibfnamefont{M.~J.} \bibnamefont{{Pierce}}},
  \bibinfo{journal}{\apj} \textbf{\bibinfo{volume}{502}}, \bibinfo{pages}{63}
  (\bibinfo{year}{1998}), \eprint{arXiv:astro-ph/9802207}.

\bibitem[{\citenamefont{{Gautret} et~al.}(2000)\citenamefont{{Gautret}, {Fort},
  and {Mellier}}}]{GauForMel00}
\bibinfo{author}{\bibfnamefont{L.}~\bibnamefont{{Gautret}}},
  \bibinfo{author}{\bibfnamefont{B.}~\bibnamefont{{Fort}}}, \bibnamefont{and}
  \bibinfo{author}{\bibfnamefont{Y.}~\bibnamefont{{Mellier}}},
  \bibinfo{journal}{\aap} \textbf{\bibinfo{volume}{353}}, \bibinfo{pages}{10}
  (\bibinfo{year}{2000}).

\bibitem[{\citenamefont{{Golse} et~al.}(2002)\citenamefont{{Golse}, {Kneib},
  and {Soucail}}}]{GolKneSou02}
\bibinfo{author}{\bibfnamefont{G.}~\bibnamefont{{Golse}}},
  \bibinfo{author}{\bibfnamefont{J.-P.} \bibnamefont{{Kneib}}},
  \bibnamefont{and}
  \bibinfo{author}{\bibfnamefont{G.}~\bibnamefont{{Soucail}}},
  \bibinfo{journal}{\aap} \textbf{\bibinfo{volume}{387}}, \bibinfo{pages}{788}
  (\bibinfo{year}{2002}), \eprint{astro-ph/0103500}.

\bibitem[{\citenamefont{{Sereno}}(2002)}]{Ser02}
\bibinfo{author}{\bibfnamefont{M.}~\bibnamefont{{Sereno}}},
  \bibinfo{journal}{\aap} \textbf{\bibinfo{volume}{393}}, \bibinfo{pages}{757}
  (\bibinfo{year}{2002}), \eprint{arXiv:astro-ph/0209210}.

\bibitem[{\citenamefont{Jain and Taylor}(2003)}]{JaiTay03}
\bibinfo{author}{\bibfnamefont{B.}~\bibnamefont{Jain}} \bibnamefont{and}
  \bibinfo{author}{\bibfnamefont{A.}~\bibnamefont{Taylor}},
  \bibinfo{journal}{Phys. Rev. Lett.} \textbf{\bibinfo{volume}{91}},
  \bibinfo{pages}{141302} (\bibinfo{year}{2003}), \eprint{astro-ph/0306046}.

\bibitem[{\citenamefont{Bernstein and Jain}(2004)}]{BerJai04}
\bibinfo{author}{\bibfnamefont{G.~M.} \bibnamefont{Bernstein}}
  \bibnamefont{and} \bibinfo{author}{\bibfnamefont{B.}~\bibnamefont{Jain}},
  \bibinfo{journal}{Astrophys. J.} \textbf{\bibinfo{volume}{600}},
  \bibinfo{pages}{17} (\bibinfo{year}{2004}), \eprint{astro-ph/0309332}.

\bibitem[{\citenamefont{{Hu} and {Jain}}(2004)}]{HuJai03}
\bibinfo{author}{\bibfnamefont{W.}~\bibnamefont{{Hu}}} \bibnamefont{and}
  \bibinfo{author}{\bibfnamefont{B.}~\bibnamefont{{Jain}}},
  \bibinfo{journal}{\prd} \textbf{\bibinfo{volume}{70}},
  \bibinfo{pages}{043009} (\bibinfo{year}{2004}), \eprint{astro-ph/0312395}.

\bibitem[{\citenamefont{Zhang et~al.}(2005)\citenamefont{Zhang, Hui, and
  Stebbins}}]{ZhaHuiSte05}
\bibinfo{author}{\bibfnamefont{J.}~\bibnamefont{Zhang}},
  \bibinfo{author}{\bibfnamefont{L.}~\bibnamefont{Hui}}, \bibnamefont{and}
  \bibinfo{author}{\bibfnamefont{A.}~\bibnamefont{Stebbins}},
  \bibinfo{journal}{Astrophys. J.} \textbf{\bibinfo{volume}{635}},
  \bibinfo{pages}{806} (\bibinfo{year}{2005}), \eprint{astro-ph/0312348}.

\bibitem[{\citenamefont{Johnston et~al.}(2007)}]{Johetal05}
\bibinfo{author}{\bibfnamefont{D.~E.} \bibnamefont{Johnston}}
  \bibnamefont{et~al.}, \bibinfo{journal}{Astrophys. J.}
  \textbf{\bibinfo{volume}{656}}, \bibinfo{pages}{27} (\bibinfo{year}{2007}),
  \eprint{astro-ph/0507467}.

\bibitem[{\citenamefont{{Hu} et~al.}(2007)\citenamefont{{Hu}, {DeDeo}, and
  {Vale}}}]{HuDeDVal06}
\bibinfo{author}{\bibfnamefont{W.}~\bibnamefont{{Hu}}},
  \bibinfo{author}{\bibfnamefont{S.}~\bibnamefont{{DeDeo}}}, \bibnamefont{and}
  \bibinfo{author}{\bibfnamefont{C.}~\bibnamefont{{Vale}}},
  \bibinfo{journal}{New J. of Phys.} \textbf{\bibinfo{volume}{\rm in press}},
  \bibinfo{pages}{astro} (\bibinfo{year}{2007}), \eprint{astro-ph/0701276}.

\bibitem[{\citenamefont{{Seljak} and {Zaldarriaga}}(2000)}]{SelZal00}
\bibinfo{author}{\bibfnamefont{U.}~\bibnamefont{{Seljak}}} \bibnamefont{and}
  \bibinfo{author}{\bibfnamefont{M.}~\bibnamefont{{Zaldarriaga}}},
  \bibinfo{journal}{\apj} \textbf{\bibinfo{volume}{538}}, \bibinfo{pages}{57}
  (\bibinfo{year}{2000}), \eprint{astro-ph/9907254}.

\bibitem[{\citenamefont{{Hu}}(2001)}]{Hu01b}
\bibinfo{author}{\bibfnamefont{W.}~\bibnamefont{{Hu}}},
  \bibinfo{journal}{\apjl} \textbf{\bibinfo{volume}{557}}, \bibinfo{pages}{L79}
  (\bibinfo{year}{2001}), \eprint{astro-ph/0105424}.

\bibitem[{\citenamefont{Hirata and Seljak}(2003)}]{HirSel02}
\bibinfo{author}{\bibfnamefont{C.~M.} \bibnamefont{Hirata}} \bibnamefont{and}
  \bibinfo{author}{\bibfnamefont{U.}~\bibnamefont{Seljak}},
  \bibinfo{journal}{Phys. Rev.} \textbf{\bibinfo{volume}{D67}},
  \bibinfo{pages}{043001} (\bibinfo{year}{2003}), \eprint{astro-ph/0209489}.

\bibitem[{\citenamefont{{Vale} et~al.}(2004)\citenamefont{{Vale}, {Amblard},
  and {White}}}]{ValAmbWhi04}
\bibinfo{author}{\bibfnamefont{C.}~\bibnamefont{{Vale}}},
  \bibinfo{author}{\bibfnamefont{A.}~\bibnamefont{{Amblard}}},
  \bibnamefont{and} \bibinfo{author}{\bibfnamefont{M.}~\bibnamefont{{White}}},
  \bibinfo{journal}{New Astronomy} \textbf{\bibinfo{volume}{10}},
  \bibinfo{pages}{1} (\bibinfo{year}{2004}), \eprint{astro-ph/0402004}.

\bibitem[{\citenamefont{Maturi et~al.}(2005)\citenamefont{Maturi, Bartelmann,
  Meneghetti, and Moscardini}}]{MatBarMenMos05}
\bibinfo{author}{\bibfnamefont{M.}~\bibnamefont{Maturi}},
  \bibinfo{author}{\bibfnamefont{M.}~\bibnamefont{Bartelmann}},
  \bibinfo{author}{\bibfnamefont{M.}~\bibnamefont{Meneghetti}},
  \bibnamefont{and}
  \bibinfo{author}{\bibfnamefont{L.}~\bibnamefont{Moscardini}},
  \bibinfo{journal}{Astron. Astrophys.} \textbf{\bibinfo{volume}{436}},
  \bibinfo{pages}{37} (\bibinfo{year}{2005}), \eprint{astro-ph/0408064}.

\bibitem[{\citenamefont{{Lewis} and {King}}(2006)}]{LewKin06}
\bibinfo{author}{\bibfnamefont{A.}~\bibnamefont{{Lewis}}} \bibnamefont{and}
  \bibinfo{author}{\bibfnamefont{L.}~\bibnamefont{{King}}},
  \bibinfo{journal}{\prd} \textbf{\bibinfo{volume}{73}},
  \bibinfo{pages}{063006} (\bibinfo{year}{2006}), \eprint{astro-ph/0512104}.

\bibitem[{\citenamefont{Metzler et~al.}(1999)\citenamefont{Metzler, White,
  Norman, and Loken}}]{Metetal99}
\bibinfo{author}{\bibfnamefont{C.~A.} \bibnamefont{Metzler}},
  \bibinfo{author}{\bibfnamefont{M.~J.} \bibnamefont{White}},
  \bibinfo{author}{\bibfnamefont{M.}~\bibnamefont{Norman}}, \bibnamefont{and}
  \bibinfo{author}{\bibfnamefont{C.}~\bibnamefont{Loken}},
  \bibinfo{journal}{Astrophys. J.} \textbf{\bibinfo{volume}{520}},
  \bibinfo{pages}{L9} (\bibinfo{year}{1999}), \eprint{astro-ph/9904156}.

\bibitem[{\citenamefont{{Hu} and {Okamoto}}(2002)}]{HuOka01}
\bibinfo{author}{\bibfnamefont{W.}~\bibnamefont{{Hu}}} \bibnamefont{and}
  \bibinfo{author}{\bibfnamefont{T.}~\bibnamefont{{Okamoto}}},
  \bibinfo{journal}{\apj} \textbf{\bibinfo{volume}{574}}, \bibinfo{pages}{566}
  (\bibinfo{year}{2002}), \eprint{astro-ph/0111606}.

\bibitem[{\citenamefont{{Hu} et~al.}(2006)\citenamefont{{Hu}, {Huterer}, and
  {Smith}}}]{HuHutSmi06}
\bibinfo{author}{\bibfnamefont{W.}~\bibnamefont{{Hu}}},
  \bibinfo{author}{\bibfnamefont{D.}~\bibnamefont{{Huterer}}},
  \bibnamefont{and} \bibinfo{author}{\bibfnamefont{K.~M.}
  \bibnamefont{{Smith}}}, \bibinfo{journal}{\ApJL}
  \textbf{\bibinfo{volume}{650}}, \bibinfo{pages}{L13} (\bibinfo{year}{2006}),
  \eprint{astro-ph/0607316}.

\bibitem[{\citenamefont{Aldering et~al.}(2004)}]{SNAP}
\bibinfo{author}{\bibfnamefont{G.}~\bibnamefont{Aldering}}
  \bibnamefont{et~al.}, \bibinfo{journal}{astro-ph/0405232}
  (\bibinfo{year}{2004}).

\bibitem[{\citenamefont{Huterer and Turner}(2001)}]{hut_turner}
\bibinfo{author}{\bibfnamefont{D.}~\bibnamefont{Huterer}} \bibnamefont{and}
  \bibinfo{author}{\bibfnamefont{M.~S.} \bibnamefont{Turner}},
  \bibinfo{journal}{Physical Review D} \textbf{\bibinfo{volume}{64}},
  \bibinfo{pages}{123527} (\bibinfo{year}{2001}),
  \urlprefix\url{astro-ph/0012510}.

\bibitem[{\citenamefont{{Vale} and {Hu}}(2007)}]{ValHu07}
\bibinfo{author}{\bibfnamefont{C.}~\bibnamefont{{Vale}}} \bibnamefont{and}
  \bibinfo{author}{\bibfnamefont{W.}~\bibnamefont{{Hu}}},
  \bibinfo{journal}{{\rm in preparation}}  (\bibinfo{year}{2007}).

\end{thebibliography}

\end{document}